\documentstyle[preprint,floats,aps,epsf]{revtex}

\begin{document}
\title{Superconductivity from doping a spin liquid insulator: a simple
one-dimensional example}
\author{Michele Fabrizio}
\address{Istituto Nazionale di Fisica della Materia, I.N.F.M.}
\address{and International School for Advanced Studies, Via Beirut 2-4,
34013 Trieste, Italy}

\maketitle
\begin{abstract}
We study the phase diagram of a one-dimensional Hubbard model
where, in addition to the standard nearest neighbor hopping $t$, we also 
include a next-to-nearest neighbor hopping $t'$. For strong enough
on-site repulsion, this model
has a transition at half filling from a magnetic insulator
with gapless spin excitations at small $t'/t$ to a dimerized insulator 
with a spin gap at larger $t'/t$. We show that upon doping this model
exhibits quite interesting features, which include the presence of
a metallic phase with a spin gap and dominant superconducting fluctuations,
in spite of the repulsive interaction.
More interestingly, we find that this superconducting phase 
can be reached upon hole doping the magnetic insulator. 
The connections between this model and the two chain models, 
recently object of intensive investigations, are also discussed.   

\vspace{2.0cm}

\noindent
SISSA Preprint no. 30/96/CM/MB
\end{abstract}
\newpage
\section{Introduction}
\label{sec:Introduction}
The properties of correlated electrons confined to a double chain 
have recently attracted considerable attention
both from the theoretical and experimental point of views.

The theoretical analyses have been mainly focused on simplified
models as two chains of
electrons interacting via a short range repulsion (e.g. the Hubbard model)
and coupled by
a transverse hopping $t_\perp$, or two $t-J$ chains coupled both by 
a transverse hopping and by a transverse exchange $J_\perp$. 
At half filling both models are equivalent to two coupled Heisenberg 
chains whose ground state has been found to be a spin liquid insulator with 
a gap in the excitation spectrum for arbitrary 
transverse coupling\cite{Strong,Fisher,tJ}.
Away from half-filling 
both models describe a metal which however maintains a finite gap for the 
spin excitations. This behavior suggests the existence of electron pairs
which is confirmed by the evidence that the dominant fluctuations
describe $4k_F$ density waves and interchain-pairing 
fluctuations\cite{Fisher,tJ,io,altri,Nagaosa,Schulz,Scalapino}.
The latter are expected to dominate for weak repulsion and sufficiently
away from half filling or, in the $t-J$ ladders, for strong $J$.

From the experimental point of view, recent measurements on ladders 
compounds like SrCu$_2$O$_3$\cite{Sr} and
(VO)$_2$P$_2$O$_7$\cite{VO} confirmed the theoretical prediction of a spin gap
at half-filling. The transition upon doping from the spin-liquid
insulator to the metal with a spin gap has also been 
verified experimentally
in the Sr doped LaCuO$_{2.5}$\cite{La}. Unfortunately no superconducting 
transition seems to occur down to 5K, which is however not in contrast with 
the theoretical predictions (it would imply either that the doping is
still low or that the interaction is too strong).

An important message which in our opinion arises from all the 
theoretical analyses 
of the two chain models and which is the subject of the present work,
is that doping a 1D spin liquid may indeed 
result in superconductivity also in the presence of repulsive interaction.
The goal of this paper is to show that this feature is shared not only by
two chain models but also by a wider class of 1D models
which do describe a spin liquid insulator at half filling.  

Among the spin models which are known to exhibit a spin-gap in the
excitation spectrum, a very simple and well studied model is the
spin-1/2 Heisenberg chain with an additional next-to-nearest neighbor exchange
\begin{equation}
\hat{H}_{JJ'} = J\sum_{i=1}^L \vec{S}_i \cdot \vec{S}_{i+1} 
+ J'\sum_{i=1}^L \vec{S}_i \cdot \vec{S}_{i+2}.
\label{HJJ'}
\end{equation}
If $J'=0$ this model is the well known Heisenberg model\cite{Haldanebis}, 
which is characterized by gapless excitations and power-law decay
of the spin correlations. If $J'=J/2$ the ground state is
exactly known\cite{Majumdar} and consists of a product of
singlets among nearest neighboring sites (dimerized state). 
There are two of these states,
which are related among each other by the translation of one lattice constant.
A finite energy gap exists between these two degenerate states and the
first excited ones\cite{Shastry}. The transition upon increasing $J'$ 
from the gapless regime to the gapped dimerized state was studied 
using bosonization by Haldane\cite{Haldane}, who predicted the transition 
to occur at $J'\simeq J/6$. Successively, Nomura and Okamoto\cite{Nomura}    
performed a detailed numerical investigation of the model and 
estimated a larger transition value of $J'\simeq J/4$. 

A model of interacting electrons which in a particular limit reproduces
the spin-model (\ref{HJJ'}) is the Hubbard model with
an additional next-to-nearest neighbor hopping ($t-t'-U$ model), 
described by the Hamiltonian
\begin{equation}
\hat{H} = -t\sum_{\sigma=\uparrow,\downarrow}\sum_{i=1}^L
\left(c^\dagger_{i\sigma}c^{\phantom{\dagger}}_{i+1\sigma} + H.c.\right)
+t'\sum_{\sigma=\uparrow,\downarrow}\sum_{i=1}^L
\left(c^\dagger_{i\sigma}c^{\phantom{\dagger}}_{i+2\sigma} + H.c.\right)
+ U\sum_{i=1}^L n_{i\uparrow}n_{i\downarrow},
\label{H}
\end{equation}
where $c_{i\sigma}$ annihilates a spin $\sigma$ electron at site
$i$ and $n_{i\sigma}=c^\dagger_{i\sigma}c^{\phantom{\dagger}}_{i\sigma}$.
At half-filling and for $U\gg (t,t')$ this model indeed maps onto (\ref{HJJ'})
with $J=4t^2/U$ and $J'=4t'^2/U$, and therefore it is a good
candidate to study the properties upon doping of a spin liquid state.

In this paper we study the phase diagram of (\ref{H}) by making use
of weak coupling Renormalization Group (RG) and bosonization.
We will show that, as a function of the parameters (electron density $n$, 
$U/t$ and $t'/t$), the phase diagram is surprisingly rich. In particular
we find that also in this simple case superconductivity may arise
from doping the spin liquid insulator, even though the electron-electron
interaction is repulsive. Moreover for $t/4<t'<t/2$ we find 
a transition upon doping from a magnetic insulator at half filling
to a metal with dominant spin and charge density wave fluctuations and
finally to a superconductor (for small $U/t$) or a metal with a
spin gap and dominant dimer wave fluctuations (at larger $U/t$). 
Although the model is purely one dimensional, this behavior is quite
suggestive especially for its similarity to the phase diagram of HTc 
superconductors.
%\begin{figure}
%\centerline{\epsfxsize=3.0in\epsfysize=4.0in\epsfbox{Fig1.ps}}
%\vspace{-2.0cm}
%\caption{{\footnotesize Energy dispersion relation of the $t-t'-U$ model for
%$t'<t/4$.}}
%\label{Fig1}
%\end{figure} 

\section{The model}
As stated in the Introduction, we are going to study the model
described by the Hamiltonian (\ref{H}) which, 
in the absence of interaction, has the following energy dispersion relation
\begin{equation}
\epsilon(k)= -2t\cos k + 2t'\cos 2k.
\label{ek}
\end{equation} 
Notice that the model has particle-hole symmetry if, at the same time,
$t'\to -t'$. Let us first analyze the dispersion relation (\ref{ek}) which
is the starting point of our perturbative analysis. 

If \underline{$t'<t/4$} the band minimum is at $k=0$ (see Fig.1).
The model is then a simple
one-band model and if the interaction $U$ is turned on we expect a
behavior qualitatively similar to the standard Hubbard model 
($t'=0$). We are not going to discuss this case in much detail,
since its behavior is very well known\cite{Hubbard}.

If \underline{$t'>t/4$} the band minima $\pm k_{min}$ move away from $k=0$ 
(which turns into a band maximum) and satisfy
the relation 
\[
\cos k_{min} = \frac{t}{4t'}.
\]
In this case two different situations may occur (see Fig.2):
\begin{itemize}
\item[(1)] if the density is such that the chemical potential
is bigger than $\epsilon(0)=-2t+2t'$, the model at low energy is effectively
a one-band model, for which the previous conclusions for
the case $t'<t/4$ apply;
\item[(2)] if, on the contrary, the chemical potential is smaller
than $\epsilon(0)$, there are four Fermi points ($\pm k_{F1}$ and
$\pm k_{F2}$), thus the model at low energy behaves as a two-band model. 
\end{itemize}
%\begin{figure}
%\centerline{\epsfxsize=3.0in\epsfysize=4.0in\epsfbox{Fig2.ps}}
%\vspace{-2.0cm}
%\caption{{\footnotesize Energy dispersion relation of the $t-t'-U$ model
%for $t'>t/4$. Also drawn are the chemical potentials corresponding
%to two different fillings:
%$\epsilon_F^{(1)}$ refers to the case when only one band is
%involved at low energy while  
%$\epsilon_F^{(2)}$ refers to the case when two bands are involved.}}
%\label{Fig2}
%\end{figure}
At half filling this implies that:
\begin{itemize}
\item if \underline{$t'<t/2$} there are only two Fermi points
$\pm k_F = \pm \pi/2$. There is therefore a simple Umklapp 
scattering since $4k_F=2\pi$, exactly like in the standard Hubbard model; 
\item if \underline{$t'>t/2$} there are four Fermi points (see Fig.2) 
satisfying
the relation $2k_{F2}-2k_{F1}=\pi$. In this case, as we are going to discuss
in the following Section, there is only a higher order Umklapp
which involves four-electron scattering at the 
Fermi surface, since $4k_{F2}-4k_{F1}=2\pi$.
\end{itemize} 

If $U\not=0$ and one is interested in the low energy behavior,
a standard approach for a 1D system 
is to linearize the band around the Fermi points:
$\epsilon(k)= \pm v_F (k\mp k_F)$  (see Fig.3a) if 
there are only two Fermi points,
while $\epsilon_1(k)= \mp v_{F1} (k\mp k_{F1})$ and 
$\epsilon_2(k)= \pm v_{F2} (k\mp k_{F2})$ if four Fermi points
are involved (see Fig.3b). The linearization is assumed to be valid
only within some cutoff range of width $\Lambda$. 
%\begin{figure}
%\centerline{\epsfbox{Fig3.ps}}
%\caption{{\footnotesize The effective low energy models: 
%(a) simple one-band model;
%(b) two-band model.}}
%\label{Fig3}
%\end{figure}
The interaction $U$ causes scattering among these
Fermi points. These scattering processes are relevant since they generate
in perturbation theory logarithmic singularities (usual Cooper's singularity
in the particle-particle channel and additional singularities in the
particle-hole channels due to the nesting property of a 1D Fermi surface).
A standard way to cope with such a log-singularities is the 
weak coupling Renormalization Group (RG), which, along with the
bosonization technique, is a very powerful tool in 1D.
Although our analysis will make use of these techniques, 
we are not going to introduce them since there exist a wide number of
articles where they have been intensively 
discussed\cite{Solyom,bosonization,twoband}.
For the two-band model, in particular, I will closely follow the analysis
of Ref.\onlinecite{io}. In this reference a two-band model resulting from
a two-coupled chain model was analyzed both via bosonization and RG.
The only difference with Ref.\onlinecite{io} is that the inner bands 
(which is $\epsilon_1(k)$ in the present case and the anti-bonding band
with transverse momentum $k_\perp=\pi$ in the two-chain model) have
opposite slopes in the two cases. The correct mapping between the two
models is therefore\cite{io}
\begin{eqnarray*}
\pm k_{F1} \;\;&\mapsto&\;\; \mp k^\pi_F, \\
\pm k_{F2} \;\;&\mapsto&\;\; \pm k^0_F. 
\end{eqnarray*}
On provision that the previous mapping is performed,
the perturbation expansion of the $t-t'-U$ two-band model
and of the two-chain Hubbard model is exactly the same at low energy
(apart from an important difference at half filling, see next Section).
Therefore we can simply borrow all the results which have been obtained
for the two-chain models and use them for the present case. 
This is what we are going to do in the following Sections\cite{notaa}. 

Already at this stage it is apparent that
the behavior of the two-chain models is similar to that of a single chain 
$t-t'-U$ model, and
that the feature which makes the two class of models equivalent is the
presence of four Fermi points in some parameter range.

\section{The model at half-filling}

If the density corresponds to one electron per site, two cases occur, as 
previously discussed.

\subsection{$t'<t/2$}
If $t'<t/2$, the low energy model is a one band model with 
Fermi momenta $\pm \pi/2$. There is a relevant Umklapp which makes
the system an insulator. However the spin excitations are gapless and,
as a consequence, the spin correlations have a power law decay at large 
distance. The model, for what it concerns the spin degrees of freedom,
behaves exactly like a Heisenberg model\cite{Hubbard,Haldanebis}. 

\subsection{$t'>t/2$}
If $t'>t/2$, the effective model is a two band model. 
It is therefore worthwhile to start with a broad outline of the behavior of
such a two-band model in 1D. 

Without Umklapp terms, two different phases exist depending on the ratio of the
Fermi velocities $v_{F1}/v_{F2}$ and $U/t$. If $t'\simeq t/2$, the
Fermi velocity of the inner band $v_{F1}\ll v_{F2}$. 
In this case RG predicts\cite{io,Fisher} that the model is a
metal with four gapless excitations (two spin and two charge
sound modes). The properties of the ground state can be inferred from the 
correlation functions which have the slowest decay at large distances.
In this case these correlation functions describe spin and
charge density waves at the incommensurate momenta
$2k_{F2}$ and $2k_{F1}$. By increasing $t'$ also $v_{F1}/v_{F2}$  
increases and at a critical $t'_c$ a transition to a different phase
occurs\cite{io,Fisher,notatc}. In this new phase the model has    
a gap for the spin excitations, and a single gapless charge mode
which corresponds to the ordinary zero sound\cite{io,Fisher,Schulz}.
There are two competing correlation functions which have the slowest asymptotic
decay. One is, in the two-chain language, the $4k_F$ charge density 
wave\cite{Nagaosa,Fisher,Schulz}.
In the language appropriate to the $t-t'-U$ model this 
function translates into the Dimer Wave (DW) correlation function
which decays at large distances like:
\begin{equation}
\chi_{DW}(x) = \langle O_{DW}(x)O_{DW}(0)\rangle
\sim \frac{cos\left[2(k_{F2}-k_{F1})x\right]}{x^{2K}} =
\frac{cos(\pi x)}{x^{2K}},
\label{DWcorr}
\end{equation}
the last equivalence being true only at half filling and
\begin{equation}
O_{DW}(x) = S^+(x)S^-(x+a) - S^+(x-a)S^-(x)
\label{DW}
\end{equation}
being the dimer order parameter\cite{Haldane}, with $a$ the lattice
constant.

The other competing correlation function is what in the two-chain
language has been identified as a kind of $d$-wave superconducting
correlation function (SC)\cite{io}. In the $t-t'-U$ model 
\begin{equation}
\chi_{SC}(x) = \langle \Delta(x) \Delta^\dagger(0) \rangle \sim
\frac{1}{x^{1/2K}},
\label{SCcorr}
\end{equation}
where 
\begin{equation} 
\Delta(x) = \sum_{p=\pm} \psi_{pk_{F1}\uparrow}(x)
\psi_{-pk_{F1}\downarrow}(x)  
- \psi_{pk_{F2}\uparrow}(x)\psi_{-pk_{F2}\downarrow}(x).
\label{SC}
\end{equation}
The Fermi operators $\psi$'s in (\ref{SC}) are defined 
around each Fermi point, i.e.
\[
\psi_{pk_{Fi}\sigma}(x) \sim {\rm e}^{ipk_{Fi}x}
\sum_{|k|<\Lambda} {\rm e}^{ikx} c_{pk_{Fi}+k,\sigma},
\]
where $i=1,2$. Notice that the existence of a spin gap already 
signals some kind of electron pairing. Due to the repulsive nature of 
the interaction, the pair wave function should have a minimum whenever the two 
electrons approach each other. This is accomplished by the minus sign in the 
expression of the pair operator $\Delta(x)$, Eq.(\ref{SC}), which 
in turns shows the importance of 
having more than two Fermi points at disposal. However, the existence
of electron pairs does not necessarily imply dominant superconductivity.
This depends upon the pair-pair interaction which in turn determines
the value of the parameter $K$.   

From Eqs.(\ref{DWcorr})-(\ref{SCcorr}), we see in fact that
if $K>1/2$ the pairing fluctuations indeed dominate
over the DW fluctuations, while the opposite occurs if $K<1/2$.
According to bosonization\cite{bosonization,Haldanebis}, 
$K$ is related to the charge
compressibility. In particular
\begin{equation}
\frac{1}{K}=\frac{4L}{\pi v_\rho} \frac{\partial^2 E}{\partial N^2},
\label{K}
\end{equation} 
where $L$ is the length of the chain, $E$ the ground state energy,
$N$ the electron number and $v_\rho$ the velocity of the charge zero
sound. The latter can be determined numerically by calculating
the energy gap between the ground state (for closed shells at total
momentum $P=0$) and the first excited state at total momentum
$P=2\pi/L$
\[
E(P=2\pi/L) - E(P=0) = \frac{2\pi}{L} v_\rho.
\]
The larger the electron-electron repulsion, the smaller the
compressibility and consequently $K$, and the more unlikely is the
dominance of superconductivity.

Since we are at half filling, we have also to take into account  
Umklapp scattering. In this case there is only one higher order Umklapp
process, which involves a four-electron scattering at the
Fermi surface (see Fig.4). From dimensional analysis it turns
out that this Umklapp is relevant if $K<1/2$. 
In this case the zero sound mode acquires a gap and the
model becomes insulating. 
%\begin{figure}
%\centerline{\epsfbox{Fig4.ps}}
%\caption{{\footnotesize The relevant Umklapp scattering when the 
%Fermi surface is made by four Fermi points.}}
%\label{Fig4}
%\end{figure}
Having discussed the possible phases of the 
$t-t'-U$ model when the Fermi surface has two branches, 
let us study in detail their occurrence at half filling.

If $t'>t'_c$ and $K<1/2$, the Umklapp is relevant and therefore 
the model is insulating with a gap in the whole excitation spectrum and 
a finite average value of the dimer order parameter Eq.(\ref{DW})
\[
\langle O_{DW}(x) \rangle = (-1)^x \cdot const.
\]
This insulating phase certainly occurs for $U\gg (t,t')$ when
the mapping to the spin model (\ref{HJJ'}) is justified.
On the other hand, for very small $U$, the parameter $K$ can be evaluated 
by perturbation
theory and it turns out to be close to one, modulo corrections of order $U$.
Therefore, provided perturbation theory is valid, $K>1/2$ for
$U\ll t$, which implies that the Umklapp scattering is irrelevant
and the model is metallic with the dominant superconducting fluctuations
Eq.(\ref{SCcorr}). Consequently we expect a transition at a 
finite $U=U_c$ from a metal with superconducting correlations 
\underline{directly} to an insulator with a dimer order.
In Fig.6 we have drawn a qualitative phase diagram for $t'>t'_c>t/2$.
At half filling, $n=1$ in the figure, there is a critical $U$
which separates the insulating regime (the bold line in the figure
which we label DI, meaning a dimer insulator)
at larger $U$ from the metal with superconducting fluctuations
at smaller $U$ (which we label SC). 

If $t/2<t'<t'_c$, we still expect a metal-to-insulator transition
at a finite $U$, but this time the metal has no spin gap and
shows dominant density wave fluctuations. The properties of the
insulating phase into which the above metal transforms at large $U$
can not be simply deduced by means of RG, whose validity is doubtful at finite
$U$. However, we tend to believe\cite{nota} that this insulator should have
the same properties of the dimer insulator which occurs for $t'>t'_c$. 

Notice that the behavior of the $t-t'-U$ model at half filling is different 
from the behavior of the two chain models also at half filling. 
There, the Umklapp term is a two-electron scattering process and is relevant 
for any $K<1$, which implies that the model is an insulator for 
any $U\not=0$\cite{Fisher}. 
   
To conclude this section, we like to point out that, according to our 
analysis, the transition at large $U$ between the insulating 
phase with power lay decay of the spin correlations and the dimer 
insulator with a spin gap is predicted to occur at $t'\simeq t/2$ or, 
in terms of the
exchange couplings, at $J'\simeq J/4$. 
\underline{This is exactly the value found
by numerical investigation of the spin model (\ref{HJJ'}) in 
Ref.\onlinecite{Nomura}}. This coincidence, which might well be accidental,
is quite surprising, since our prediction is based simply
on {\sl band-structure arguments} (modification of the Fermi surface).
In fact, one would rather believe that band structure details are irrelevant
for electron systems in the strong correlation limit where the
interaction is much larger than the bandwidth, which is the case of
the model (\ref{HJJ'}).

\section{The model away from half filling}

We have seen that already at half filling the $t-t'-U$ model shows
the unusual property of a transition at a critical $U$  
from a metal with superconducting 
fluctuations to an insulator, when $t'>t/2$. Away from half filling
the behavior is even more interesting, for smaller and larger $t'$.
Let us consider in detail the various possible scenarios.

\subsection{$t'<t/4$}
In this case the model is for any filling
always an effective one-band model, where nothing special occurs.
Whether we dope with holes or electrons, as soon as we move away from
half-filling the charge gap closes and the system becomes a metal
with gapless spin and charge sound modes. The dominant fluctuations are
both spin and charge density wave fluctuations (SDW and CDW) 
at momentum $2k_F$.
In Fig.5 we have drawn the phase diagram for this case
as a function of $U/t$ and of the density $n$. The bold line at 
density $n=1$, labeled MI, identifies the 
magnetic insulator with power law decay of the spin correlations, while the 
rest of the phase diagram has been
labeled with SDW/CDW implying it is a metal with dominant density
wave fluctuations. These labels will be used with the same meaning 
also in the following cases.
%\begin{figure}
%\centerline{\epsfbox{Fig5.ps}}
%\caption{{\footnotesize Phase diagram of the $t-t'-U$ model for $t'<t/4$ as
%a function of $U/t$ and of the density $n$.}}
%\label{Fig5}
%\end{figure}

\subsection{$t'>t/2$} 
Here, it makes a difference whether we dope with holes or electrons
(the model is not particle-hole symmetric).
In Fig.6 we have drawn a qualitative phase diagram for $t'>t'_c>t/2$.
%\begin{figure}
%\centerline{\epsfbox{Fig6.ps}}
%\caption{{\footnotesize Qualitative phase diagram of the $t-t'-U$ model for 
%$t'$ sufficiently larger than $t/2$ as
%a function of $U/t$ and of the density $n$.}}
%\label{Fig6}
%\end{figure}

For hole doping ($n<1$), the effective low energy model involves always 
two bands.
Therefore we predict that for any hole doping the spin gap will survive.
As concerns the charge gap, it will immediately disappear as soon
as we move away from half filling. Therefore, as $U$ increases, we
expect a crossover from a metal with dominant superconducting fluctuations
[see Eq.(\ref{SCcorr})], which we still label in Fig.6 as SC since it is
continuously connected to the analogous state at $n=1$,
to a metal with dominant dimer wave
fluctuations [see Eq.(\ref{DWcorr})], which is labeled by DW.

At this point, it is worthwhile to discuss briefly the properties of the weakly
doped dimer insulator. If we approach half-filling with $U>U_c$
then, according to the theory of the incommensurate-to-commensurate
transitions in 1D\cite{CIT}, $K$ tends asymptotically to the
value $1/4$ and, exactly at half-filling, it jumps abruptly to
zero. With the asymptotic value $K=1/4$ valid at low doping, 
the dimer wave correlation function Eq.(\ref{DWcorr}) decays 
like $\chi_{DW}\sim (-1)^x/\sqrt{x}$, while the superconducting correlation
Eq.(\ref{SCcorr}) decays quadratically $\chi_{SC}\sim 1/x^2$. 
Notice that these power law decays are typical of the Green functions
and of the density-density correlation functions, respectively,
of a hard core Bose gas. 
The situation is opposite for the metallic phase at $U\ll t$. 
In this case, provided perturbation theory is valid,
$K\simeq 1$ and the behavior of the two correlation functions are exchanged:
$\chi_{SC}\sim 1/\sqrt{x}$ while $\chi_{DW}\sim (-1)^x/x^2$.

In the case of electron doping the situation is in general different
but for low doping where all is the same except the chemical potential
which moves up instead of down (see Fig.2). At the same time the Fermi 
velocities $v_{F1}$ and $v_{F2}$ of the two linear bands 
get more and more different (actually $v_{F1}\to 0$).
As we said in the previous Section, at a critical value of
$v_{F2}/v_{F1}$, or equivalently a critical density, $n_{c1}$ in Fig.6, 
RG predicts a transition to another phase where also the spin gap closes 
(see e.g. Appendix B in Ref.\onlinecite{io}). In this phase the model is
a metal with four gapless sound modes (two spin and two charge modes).
The dominant fluctuations describe charge and spin density waves.
We have labeled this phase in Fig.6 as SDW/CDW II, implying that
the number of gapless excitations is twice that of the phase SDW/CDW.

Finally, at a second critical doping $n_{c2}$, 
the topology of the Fermi surface
changes from a four-point to a two-point Fermi surface.
At the transition the density of states of the inner band diverges
due to a van Hove singularity. For this reason we are not able to predict
what happens exactly at the transition. According to Balents and 
Fisher\cite{Fisher}, the van Hove singularity induces again a spin
gap and therefore they expect the properties of the model to be
similar to those at low doping. On the other hand, if we assume that
the RG equations of Ref.\onlinecite{io} can be extended up to a very
large $v_{F2}/v_{F1}$ (where their validity is not fully guaranteed),    
we would rather expect that for $v_{F2}/v_{F1}\gg 1$ the two
linear bands effectively decouple. In this case the transition would be
a standard
metal-to-metal transition with a topological modification of the Fermi
surface. The van Hove singularity related to the low
(hole) doping of the inner band is not expected to play any
fundamental role, similarly to what happens to any one-band model close to
filling zero or one. Although we have no rigorous proof, we tend to believe 
that nothing special occurs at the transition (the latter scenario), rather 
than in the scenario proposed in Ref.\onlinecite{Fisher}. 
Coming back to the phase diagram, in the region where only two
Fermi points are involved the model should be metallic with still 
dominant density wave fluctuations (see Fig.6), and only two
gapless modes (therefore this phase is labeled SDW/CDW).  

In Fig.6 we have assumed that the critical density $n_{c1}$ decreases
by increasing $U$. This is true for very small $U$ where perturbation theory
is valid. For finite $U$, unfortunately, we have no reliable method to 
evaluate $n_{c1}$. It might well be that $n_{c1}\to n_{c2}$ at $U\gg (t,t')$, 
which would be more consistent with the proposed transition upon increasing
$U$ at half filling
and for $t'\simeq t/2$ from the phase CDW/SDW II to the dimer insulator DI. 
The uncertainty about the precise behavior of $n_{c1}$
as a function of $U$ is also the reason why we have prefered not to draw
even a qualitative phase diagram for $t'$ close to $t/2$.

\subsection{$t/4<t'<t/2$} 
This case is in our opinion the most interesting one 
for its surprising similarities with the behavior of HTc compounds.
We said in the previous Section that at half filling the model
is an insulator (MI in Fig.7 at $n=1$) with power law decaying spin 
correlations.
If we dope with electrons ($n>1$), the charge gap suddenly closes
and the model turns into a metal with dominant spin and charge
density wave fluctuations, for arbitrary doping (SDW/CDW in Fig.7).
%\begin{figure}
%\centerline{\epsfbox{Fig7.ps}}
%\caption{{\footnotesize Qualitative phase diagram of the $t-t'-U$ 
%model for $t/4<t'<t/2$ as
%a function of $U/t$ and of the density $n$.}}
%\label{Fig7}
%\end{figure}

If we dope with holes ($n<1$), at low doping
the Fermi surface still consists of two points, and therefore the
properties are similar to what we just described for electron doping,
i.e. the phase is a SDW/CDW.
However, as the chemical potential moves down (see Fig.2), 
the same situation encountered earlier re-appears, 
now in reverse.
At a lower critical doping ($n_{c2}$, in analogy with the
previous case) there will be a
topological modification of the Fermi surface from a two-point to a
four-point surface. The model turns from a one-band to a two-band model.
Until the Fermi velocities remain quite different ($v_{F2}/v_{F1}\gg 1$)
the system is a metal with four gapless modes (two charge and
two spin modes) and dominant density wave fluctuations (phase SDW/CDW II in 
Fig.7). 
At a critical value of $v_{F2}/v_{F1}$, or equivalently a 
critical doping $n_{c1}$, a spin gap opens and only
one charge mode (the ordinary zero sound) remains gapless. 
The dominant fluctuations are dimer waves Eq.(\ref{DW})
or pairing fluctuations, Eq.(\ref{SC}). The one which dominates depends
on the value of the parameter $K$ Eq.(\ref{K}), which in turns depends
on $U$. As we said before, if $K<1/2$ the dimer fluctuations win 
(DW phase in Fig.7), 
while for $K>1/2$ superconductivity is more relevant (SC phase in Fig.7). 
Like in the previous case, we can not establish the precise behavior 
at large $U$  
of the critical line between the CDW/SDW II phase and the SC or DW phases.
Therefore the shape of the phase diagram drawn in Fig.7 has not to be
taken too literally close to that critical line.

Notice that the appearance of the spin gap upon hole doping the magnetic 
insulator indicates that in this model the holes \underline{increase} 
the spin frustration.
In the large $U$ limit, one can map the $t-t'-U$ model onto a generalized
$t-J$ model with the Hamiltonian
\begin{eqnarray*}
\hat{H} &=& -t\sum_{\sigma=\uparrow,\downarrow}\sum_{i=1}^L
\left(c^\dagger_{i\sigma}c^{\phantom{\dagger}}_{i+1\sigma} + H.c.\right)
+t'\sum_{\sigma=\uparrow,\downarrow}\sum_{i=1}^L
\left(c^\dagger_{i\sigma}c^{\phantom{\dagger}}_{i+2\sigma} + H.c.\right)\\
& & + J\sum_{i=1}^L \vec{S}_i \cdot \vec{S}_{i+1} 
+ J'\sum_{i=1}^L \vec{S}_i \cdot \vec{S}_{i+2},
\end{eqnarray*}
defined in the reduced Hilbert space where double occupancies are forbidden. 
This model with $t'=0$ has been 
numerically investigated by Ogata, Luchini and Rice\cite{Ogata}.
They find that the hole doping effectively reduces the frustration
due to $J'$. Since we instead find an increase of frustration, 
we have to conclude that this is mainly a consequence of the 
next-to-nearest-neighbor hopping $t'$\cite{spin-charge}.

\section{Conclusions}
In this paper we have studied the phase diagram of 
a one-dimensional $t-t'-U$ model where, in addition to the on site
repulsion $U$ and nearest neighbor hopping $t$, we have included a
next-to-nearest neighbor hopping $t'$. Although very simple, 
this single chain model has the interesting property to be at half filling 
and large $U$ either an insulator with gapless spin excitations
or a dimerized insulator with a spin gap\cite{Haldane}.  
The transition between the two insulators should occur, according to
our analysis, at $t'\simeq t/2$. 
This model is therefore particularly suited to study the occurrence
of superconductivity upon doping an insulator which has a
charge gap and a spin gap. 
In fact, recent theoretical and numerical investigations of 
two coupled chain models\cite{tJ,Scalapino}, 
suggest 
that the presence of a spin gap in the insulating phase of two chains 
at half filling may lead to superconductivity upon doping and for not too
large repulsion. We have shown that precisely this behavior 
is realized here.
Moreover, the weak coupling analysis seems to suggests
that the key feature which is responsible for the spin gap and possibly for
the superconductivity is the topology of the Fermi surface in both models,
which in the interesting parameter range 
has two branches, i.e. four Fermi points.  

The present $t-t'-U$ model has additional properties which make this
model interesting in its own right. 
In particular at half-filling and for $t'$ sufficiently
larger than $t/2$, we
predict a \underline{direct} transition at a critical $U_c$ from a metal with 
dominant superconducting fluctuations at $U<U_c$ to a dimerized
insulator at $U>U_c$. 

Interestingly, for
$t/4<t'<t/2$ the phase diagram for hole doping shows some
similarities with the phase diagram of HTc compounds (see Fig.7).
At half-filling we have a magnetic insulator with power law decay of 
the spin correlations. For low doping we move to a metal with dominant
spin and charge density waves fluctuations. Above a critical
doping, a spin gap opens and the model has either dominant superconductivity
or dimer waves depending upon the strength of the on site repulsion.   

\section{Acknowledgments}
I gratefully acknowledge helpful discussions with E. Tosatti, 
C. Castellani, A. Parola and G. Santoro. This work has been partly 
supported by EEC, under contract ERB CHR XCT 940438.

\begin{figure}
\vspace{-7.0cm}
\centerline{\epsfbox{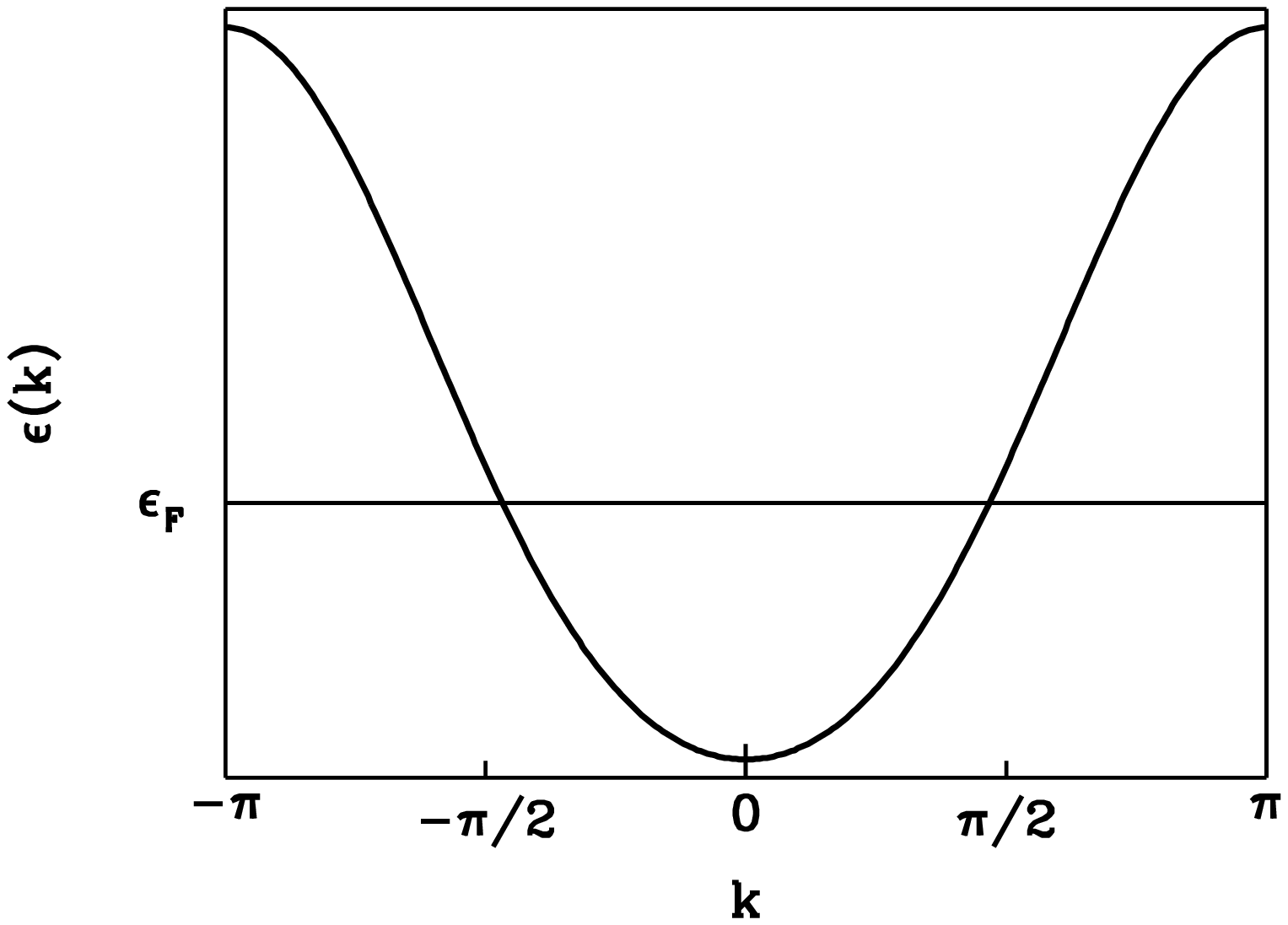}}
\vspace{-2.5cm}
\caption{ Energy dispersion relation of the $t-t'-U$ model for
$t'<t/4$.}
\label{Fig1}
\end{figure}
\begin{figure}
\centerline{\epsfbox{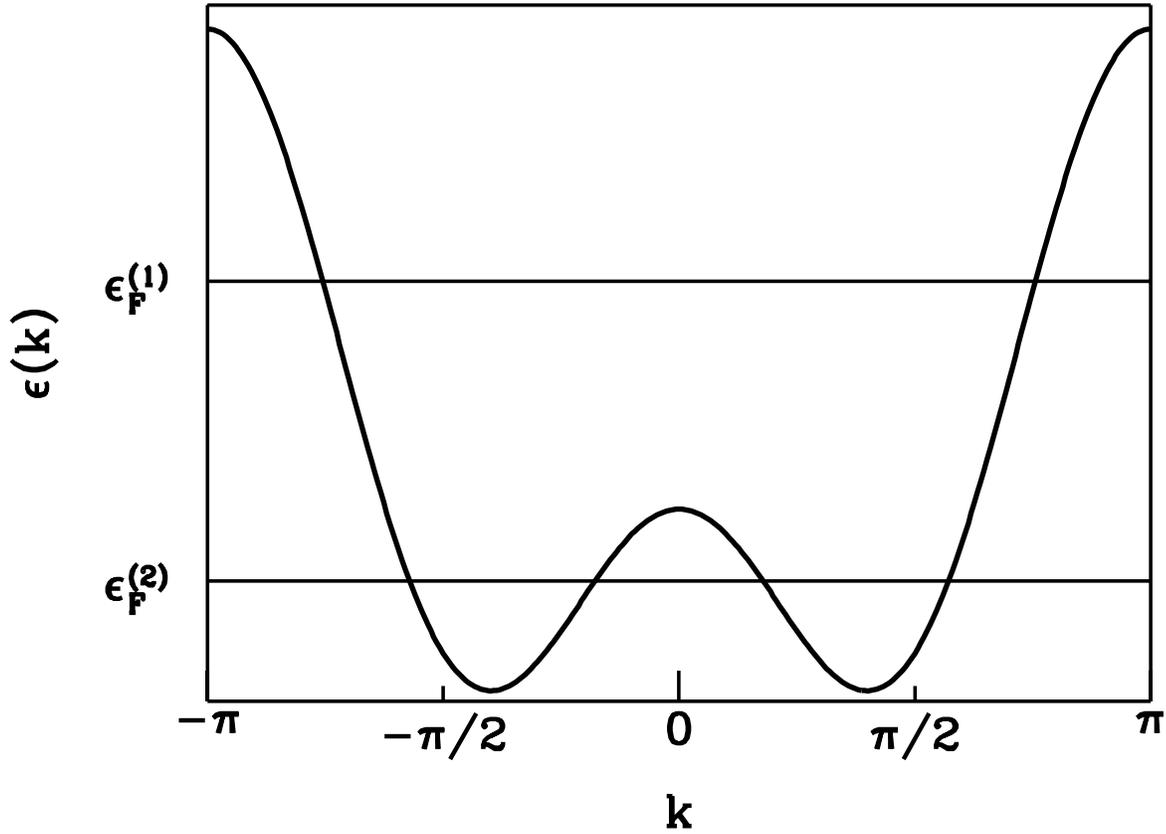}}
\vspace{-2.5cm}
\caption{ Energy dispersion relation of the $t-t'-U$ model
for $t'>t/4$. Also drawn are the chemical potentials corresponding
to two different fillings:
$\epsilon_F^{(1)}$ refers to the case when only one band is
involved at low energy while  
$\epsilon_F^{(2)}$ refers to the case when two bands are involved.}
\label{Fig2}
\end{figure}
\begin{figure}
\vspace{4.0cm}
\centerline{\epsfbox{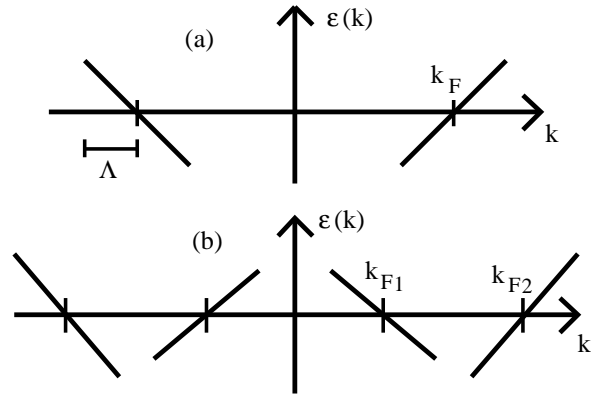}}
\caption{The effective low energy models: (a) simple one-band model;
(b) two-band model.}
\label{Fig3}
\end{figure}
\begin{figure}
\vspace{4.0cm}
\centerline{\epsfbox{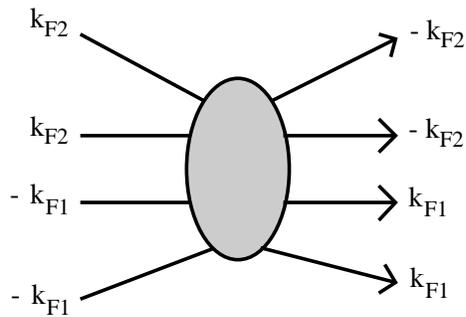}}
\caption{ The relevant Umklapp scattering when the 
Fermi surface is made by four Fermi points.}
\label{Fig4}
\end{figure}
\begin{figure}
\vspace{4.0cm}
\centerline{\epsfbox{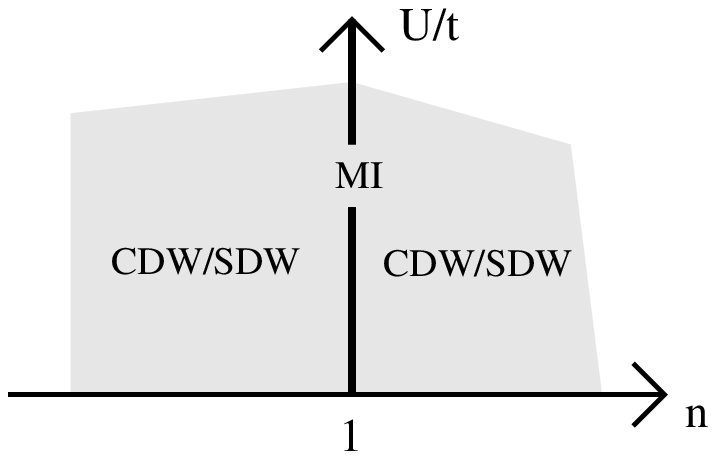}}
\caption{Phase diagram of the $t-t'-U$ model for $t'<t/4$ as
a function of $U/t$ and of the density $n$.}
\label{Fig5}
\end{figure}
\begin{figure}
\vspace{4.0cm}
\centerline{\epsfbox{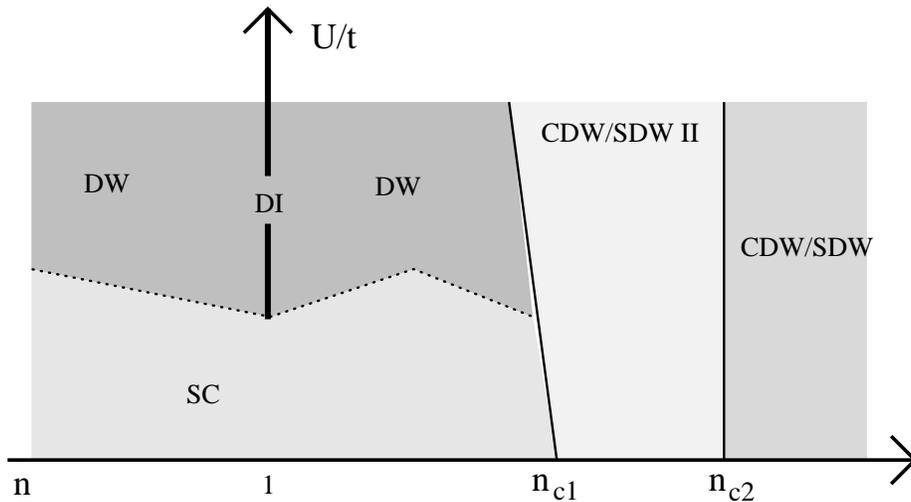}}
\caption{ Qualitative phase diagram of the $t-t'-U$ model for 
$t'$ sufficiently larger than $t/2$ as
a function of $U/t$ and of the density $n$.}
\label{Fig6}
\end{figure}
\begin{figure}
\vspace{4.0cm}
\centerline{\epsfbox{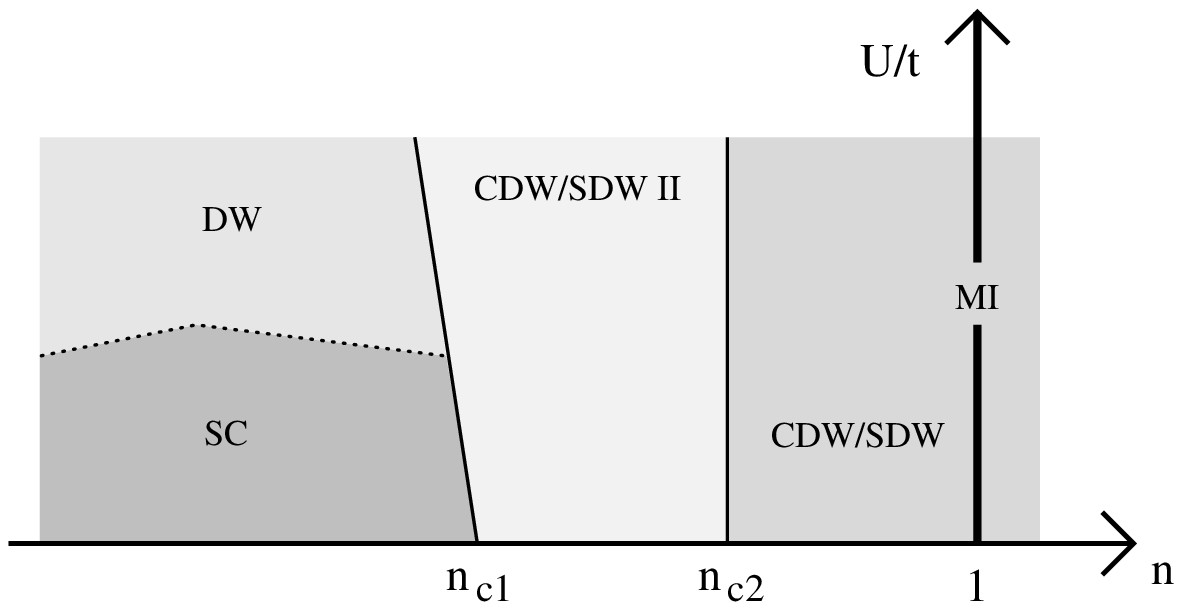}}
\caption{Qualitative phase diagram of the $t-t'-U$ 
model for $t/4<t'<t/2$ as
a function of $U/t$ and of the density $n$.}
\label{Fig7}
\end{figure}


\begin{references}
\bibitem{notaa}It is worthy to notice that in the RG approach it is possible
to renormalize also the difference between the Fermi points of the
two linearized band. However, as noticed in Ref.\onlinecite{io},
this has no effect for weak starting repulsion, where RG is valid.
For strong starting repulsion, the RG approach would predict additional phases,
which are likely to be a spurious result since weak coupling RG can not 
be applied in that regime. More refined 
analyses (see e.g. Ref.\onlinecite{Schulz}) suggest that this is indeed 
the case.
\bibitem{notatc}$t'_c$ depends in general on $U/t$. It is different from
$t/2$ when $U\to 0$, and increases upon increasing $U\ll t$. For large $U/t$
its behavior can not be obtained by means of the perturbative techniques
used in this paper.
\bibitem{nota}Our belief is based on the (not rigorous) argument that
the difference between the two Fermi velocities should be an 
irrelevant parameter in the insulating phase, and that the insulator which
we find starting from equal velocities is just the dimerized one.
\bibitem{spin-charge}Notice that the next-to-nearest 
neighbor hopping $t'$ might be expected to lead to non trivial consequences.
In fact in the presence of a finite $t'$ 
the factorization of the wave function at large $U$ 
into a charge and a spin part is not anymore valid, contrary to the
standard Hubbard model or to the $t-J-J'$ model in the
limit of small exchange couplings.
\end{references}
\end{document}